\documentclass[letterpaper, 12 pt, conference]{ieeeconf}  

\IEEEoverridecommandlockouts                              
\usepackage{hyperref}
\usepackage{amsmath}
\usepackage{amssymb}
\usepackage{graphicx}
\hypersetup{
    colorlinks=true,
    linkcolor=blue,
    filecolor=magenta,      
    urlcolor=cyan,
}

\title{\LARGE \bf
Backward Propagation
}

\author{Anurag Pallaprolu\\
ECE Department, UCSB
}

\begin{document}

\maketitle
\thispagestyle{empty}
\pagestyle{empty}

\begin{abstract}
The document covers the fundamental algorithm of backward propagation from the point of view of reconstructing the wavefield captured by a "screen" in an imaging system. Owing to a property of the Helmholtz equation, wavefields have an inherent propagation symmetry which can be exploited for image formation. In traditional presentations of this topic the link between this "reciprocity" and the reconstruction procedure are motivated only at an intuitive level. I propose to make this more rigorous by using a technique called "plane-to-plane" propagation which is also known as "beam propagation method" in more advanced settings.
\end{abstract}

\vspace{3 mm}
The backward propagation/backward projection/back-projection algorithm is a relatively old technique to form an image of a source region that has been split into finite elements, with the quality of the image depending on the granularity of the grid. This forms the basis for the Fourier reconstruction techniques encountered when studying tomography based systems such as CT, MRI, et cetera [1, 2, 3]. In order to elaborate on the method itself we will have to consider a source region which we assume is distributed with point sources at locations $\bar{r}_1, \bar{r}_2, ..., \bar{r}_N$. We can then assume that the overall source region can be represented as a convolution:

$$\mathcal{W}(\bar{r}) = \sum_i \mathcal{A}_i\delta(\bar{r} - \bar{r}_i)$$

where $\mathcal{A}_i$ is the value of the amplitude of the $i^{th}$ source. Before we proceed with the solution of this original configuration over a figurative screen, it would be nice to see the interpretation of the Dirac delta source in different contexts. The delta function of one variable $\delta_1(x)$ has the description of being a supposedly well defined function/distribution on a (compact) support which is $\infty$ at $x=0$ and $0$ elsewhere. Extending this definition to two variables, we get $z = \delta_2(x, y)$ is now a contour which is $z = 0$ everywhere except at $x = y = 0$. For three and four dimensions we have representations $\delta_3(x, y, z)$ and $\delta_4(x, y, z, t)$ defined in similar manner. While these facts may sound obvious, there is a point that is often not emphasized with regards to the higher dimensional variants: 
$$\delta_1(x) = \delta_3(x, y_0, z_0)$$

where $y_0, z_0$ are constants i.e., the 1D delta function can be seen as a projection of the 3D flavor onto a space constrained by $y=y_0, z=z_0$. This is nothing but a plane in $\mathbb{R}^3$ and one can consequently see that $\delta_1$ is a line in $\mathbb{R}^2$ and a point/sphere of infinitesimal radius in $\mathbb{R}$. Thus while imaging a dynamic patch of area in general, where $\bar{r} = (x, y, z, t) \in \mathbb{R}^4$, we could consider $\delta(x, y, z)$ as the "spike" at the origin but at only the very initial instant of time $t = 0$. Thus for a complete analysis of a dynamic source region, we will have to not only consider the "temporal" frequency $f$ which is the Fourier conjugate of $t$, but also the three "spatial" frequencies $(f_x, f_y, f_z)$ which are therefore the Fourier conjugates of $(x, y, z)$. In the most general sense, we will then have:

$$\mathcal{F}(A\delta(\bar{r})) = A\mathcal{F}(\delta(x, y, z, t)) = Ae^{j<\bar{f}, \bar{r}>}$$

So let us assume there is an infinite one dimensional "screen" aligned with the $x$-axis and let there be a coherent plane wave coming from the positive $y$-axis (originating from a point source sitting at $y = +\infty$) as shown in Figure 1.
\begin{figure}[!htb]
    \centering
    \includegraphics[scale=0.55]{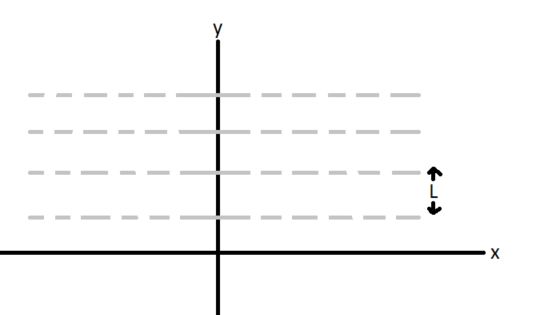}
    \caption{Screen along x-axis for source at $y = \infty$}
\end{figure}
The screen which physically consists of molecules getting excited in unison, will have a frequency of this excitation equal to $\frac{1}{L}$, and if we placed multiple screens at $y = k$ for varying $k$, we would characterize $f_y = \frac{1}{L}$ as well. The same hypothesis would extend to the acquisition of a coherent plane-wave originating from $x = \infty$ as shown in Figure 2.
\begin{figure}[!htb]
    \centering
    \includegraphics[scale=0.55]{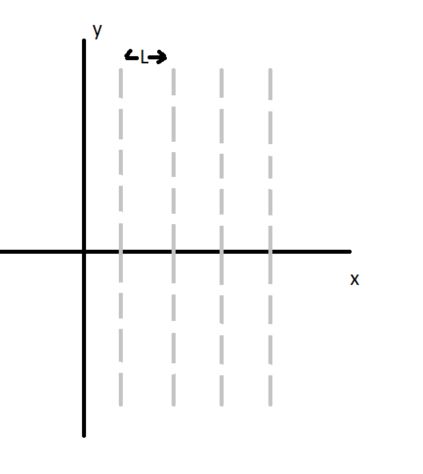}
    \caption{Screen along y-axis for source at $x = \infty$}
\end{figure}
We can evolve this aligned source setup to a superposition by placing the point source at potentially $(x, y) = (\infty, \infty)$ i.e., at far field but algined at an angle of $\frac{\pi}{4}$. We would then have the following image for the reception, and in this case we would need to have infinite screens aligned across both $x$ and $y$ axes to capture the complete wavefield information, as shown in Figure 3. 
\begin{figure}[!htb]
    \centering
    \includegraphics[scale=0.55]{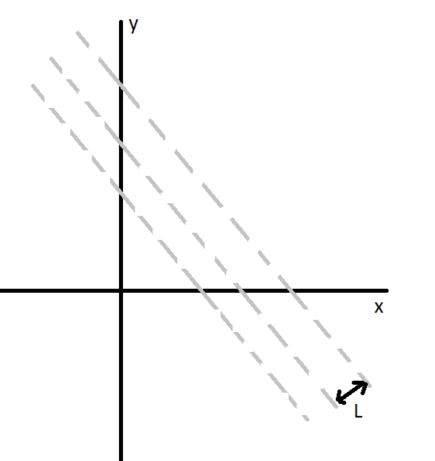}
    \caption{Screen along x and y-axes for source at $x, y = \infty, \infty$}
\end{figure}
Obviously, the magnitude of the angle made by each incoming wavefront by the positive $x$-axis is $\frac{\pi}{4}$ as well but now points on both the screens will get excited with frequencies that are not directly related to $L$. Instead, we will have $\Delta x = \Delta y = \frac{L}{sin\frac{\pi}{4}} = \sqrt{2}L$. In a more general setting, we will have:
$$\Delta x = \frac{L}{\sin\theta}$$
$$\Delta y = \frac{L}{\cos\theta}$$
Thereby, we can eliminate $\theta$ to see that:
$$\frac{1}{(\Delta x)^2} + \frac{1}{(\Delta y)^2} = \frac{1}{L^2}$$
$$\implies f_x^2 + f_y^2 = \frac{1}{L^2}$$
In a more general, three-dimensional setting, we have:
$$f_x^2 + f_y^2 + f_z^2 = \frac{1}{L^2} = \frac{f^2}{c^2}$$
where $c$ is the propagation speed of the wavefront. Thus, we see that the impact of a far field point source on screens that cover all degrees of freedom is a "shell" in the frequency domain centered at the origin with a radius of $\frac{f}{c}$. This also establishes a connection between the "spatial" and "temporal" frequencies we were discussing about earlier. In order to extend this analysis to actually reconstructing the image of the far-field object from the information captured by these "screens", we will have to analyze the propagation of the wavefield itself, which is governed by the Helmholtz partial differential equation[4]:
\\
$$(\nabla_{\bar{r}}^2 + \kappa^2)\mathcal{W}(\bar{r}) = \mathcal{S}(\bar{r})$$
\\
where $\kappa = \frac{2\pi}{L}$ is the so-called wave number, $\mathcal{W}$ is the  response of the radiation captured, and $\mathcal{S}$ is the source. When seen from a systems perspective, we could imagine that the operator $\mathcal{V} = (\nabla_{\bar{r}}^2 + \kappa^2)^{-1}$ is a "transfer" function of sorts, carrying the source configuration to the response/destination i.e., $\mathcal{W} = \mathcal{V}\mathcal{S}$. Likewise, we will have to find the impulse response of the operator $\mathcal{V}$ to get the actual transfer function, and this is done by setting the source to a delta function:
\\
$$(\nabla_{\bar{r}}^2 + \kappa^2)h(\bar{r}) = \delta(\bar{r})$$
\\
Except the neighborhood of $\bar{r} = (0, 0, 0)$ we would have the following homogeneous PDE:
\\
$$(\nabla_{\bar{r}}^2 + \kappa^2)h(\bar{r}) = 0$$
\\
Taking the Fourier transform of both sides, and recalling that the Fourier transform of the derivative operator $\frac{\partial}{\partial x_k}$ is nothing but pre-multiplying by $j\omega_k = j2\pi f_k$, we have:
\\
$$[(j2\pi f_x)^2 + (j2\pi f_y)^2 + (j2\pi f_z)^2 + \kappa^2]\mathcal{H}(\bar{f}) = 0$$
$$\implies [\frac{1}{L^2} - (f_x^2 + f_y^2 + f_z^2)]\mathcal{H}(\bar{f}) = 0$$
\\
Thus, in a rather semi-rigorous manner, we can define the solution $\mathcal{H}(\bar{f})$ as:
\\
$$\mathcal{H}(\bar{f}) = \delta_3(f_x^2 + f_y^2 + f_z^2 - \frac{1}{L^2})$$
$$\implies h(\bar{r}) = \mathcal{F}^{-1}(\mathcal{H}) = \frac{2}{L|\bar{r}|}sin(\frac{2\pi |\bar{r}|}{L})$$
\\
The amplitude term is due to the classical Hugyens-Fresnel principle which necessitates the conservation of energy with every step of the wavefront propagation (thus we would not have non-zero fields at say truly infinite radius from the point of origin). Also the result depends purely on the magnitude of the radial distance and is therefore spherically symmetric: this structure goes by the name of the Green kernel. We will therefore represent $|\bar{r}| = r$ and proceed. Using De Moivre's theorem:
\\
$$h(r) = \frac{2}{Lr}sin(\frac{2\pi r}{L}) = \frac{1}{jL r}e^{\frac{j2\pi r}{L}} - \frac{1}{jL r}e^{\frac{-j2\pi r}{L}}$$
$$\implies h(r) = \frac{1}{jL r}e^{\frac{j2\pi r}{L}} + \frac{1}{jL (-r)}e^{\frac{j2\pi (-r)}{L}}$$
\\
This form of the expression gives us the famous Helmholtz reciprocity rule. That is, $h(r) = h(-r)$ and this implies that from a localized impulse source you will have \textit{two} wavefronts expanding symmetrically, given by the two complex exponentials. Since they only differ in the phase factor, we will take only the "forward" propagator and write:
\\
$$h_G(r) = \frac{1}{jLr}e^{\frac{j2\pi r}{L}}$$
\\
This is called the Green-Born propagator and noting that the wave equation is linear in the response characteristic, we can sum over all such propagators over the imaging area to reconstruct the final image of a compound object. This fact combined with the reciprocity rule forms the theoretical basis for the backward propagation technique. To complete the picture (!), we will discuss the so-called "plane-to-plane" propagation idea. Let us assume the same generic point source at $(\infty, \infty)$ and place two screens, one at $y = 0$ and the other at a generic $y$, as shown in Figure 4.
\begin{figure}[!htb]
    \centering
    \includegraphics[scale=0.55]{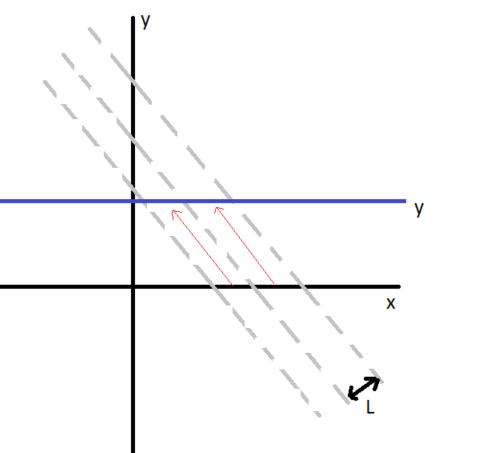}
    \caption{Screens parallel to x-axis for source at $x, y = \infty, \infty$}
\end{figure}
While the patterns of excitation on the screens at $y=0$ and generic $y$ will be the same in the frequency content, they are clearly off by a phase shift, and we can utilize the spatio-temporal condition to compute this value:
\\
$$f_y^2 + f_x^2 = \frac{1}{L^2}$$
$$\implies f_y = \sqrt{\frac{1}{L^2} - f_x^2}$$
$$\implies \frac{d\phi_y}{dy} = 2\pi\sqrt{\frac{1}{L^2} - f_x^2}$$
$$\implies \phi_y = 2\pi y\sqrt{\frac{1}{L^2} - f_x^2}$$
\\
This simple example shows that moving the screen through a patch of "source-free" region induces only a phase shift that is proportional to the distance the screen moves by. Using the reciprocity criterion, we can also argue that this is the same effect that the wavefront (represented by the dashed lines) undergoes when propagating through such a "source-free" region. This is called the plane-to-plane propagation technique. Basically, we can take the excitation recorded by the screen, phase shift each point by $\phi_y$ thereby virtually moving the screen backwards by a distance $y$ and reconstructing \textit{what would have been} the wavefront at a distance $y$ before arriving at the screen. Thus, the effect on the Fourier space representation of the screen would be an exponential phase factor multiplication, and in the three dimensional setting (i.e. our screen is 2 dimensional), we have the following relation:
\\
$$\mathcal{S}_{z = z_2}(f_x, f_y) = \mathcal{S}_{z = z_1}(f_x, f_y) \times e^{j\phi_z}$$
$$\phi_z(f_x, f_y) = 2\pi |(z_1 - z_2)|\sqrt{\frac{1}{L^2} - f_x^2 - f_y^2}$$
\\
Once again we can note that the phase shift consists of terms which depend purely on the magnitude of the movement of the screen/wavefront. Also, due to the square root, we have a constraint on the range of values the frequency vectors can take, namely
\\
$$\frac{1}{L^2} - f_x^2 - f_y^2 \geq 0$$
$$\implies f_x^2 + f_y^2 \leq \frac{1}{L^2}$$
\\
This establishes a resolution on the image being backward propagated so to speak and due to the inequality above, the system as such behaves like a low pass filter. In other words, wavefront propagation from plane to plane is nothing but a low pass filter with a bandwidth of $\frac{2}{L}$. 
\\
\\
The backward propagation technique is only but a minor modification of this idea and for this we will have to flip the sign of $\Delta z = |z_1 - z_2|$ in the exponential since we are now propagating backwards (reversing the ordinary plane-to-plane motion). We can observe that the inverse Fourier transform of $\phi(f_x, f_y, -z) = \phi_{-z}$ with respect to frequency space variables $f_x, f_y$ is nothing but
\\
$$\mathcal{F}^{-1}[exp(-j2\pi \Delta z \sqrt{\frac{1}{L^2} - f_x^2 - f_y^2})] = \frac{-1}{jLr}e^{\frac{-j2\pi r}{L}}$$
\\
which can equivalently be seen as a Green-Born propagator, moving backwards by a distance $r$. Thus, the distance space variation of the screen can be seen as a \textit{conjugation} operation with the Green kernel during the backward propagation:
\\
$$S(\bar{r}_1) = \frac{1}{jL(\Delta r)}e^{\frac{j2\pi (\Delta r)}{L}} * S(\bar{r}_2)$$
\\
and this completes the demonstration.
\section*{Acknowledgements}
I would like to thank Prof. Hua Lee and Prof. Yasamin Mostofi of the ECE Department, UCSB for their illuminating discussions and corrections. 
\section*{References}
1. Dasch, Cameron J. "\textit{One-dimensional tomography: a comparison of Abel, onion-peeling, and filtered backprojection methods.}" Applied optics 31.8 (1992): 1146-1152.
\\
\\
2. Lauritsch, G\"unter, and Wolfgang H. H\"arer. "\textit{Theoretical framework for filtered back projection in tomosynthesis.}" Medical Imaging 1998: Image Processing. Vol. 3338. International Society for Optics and Photonics, 1998.
\\
\\
3. M\"uller, Paul, Mirjam Schürmann, and Jochen Guck. "\textit{The theory of diffraction tomography}." arXiv preprint arXiv:1507.00466 (2015).
\\
\\
4. J. W. Goodman. \textit{Introduction to Fourier Optics} (2nd ed.). pp. 61, 62.
\end{document}